# Modified physics-informed neural network method based on the conservation law constraint and its prediction of optical solitons

Gang-Zhou Wu, Yin Fang, Yue-Yue Wang[*] and Chao-Qing Dai [*]

*College of Optical, Mechanical and Electrical Engineering, Zhejiang A&F University, Lin'an 311300, China*

**Abstract.** Based on conservation laws as one of the important integrable properties of nonlinear physical models, we design a modified physics-informed neural network method based on the conservation law constraint. From a global perspective, this method imposes physical constraints on the solution of nonlinear physical models by introducing the conservation law into the mean square error of the loss function to train the neural network. Using this method, we mainly study the standard nonlinear Schrödinger equation and predict various data-driven optical soliton solutions, including one-soliton, soliton molecules, two-soliton interaction, and rogue wave. In addition, based on various exact solutions, we use the modified physics-informed neural network method based on the conservation law constraint to predict the dispersion and nonlinear coefficients of the standard nonlinear Schrödinger equation. Compared with the traditional physics-informed neural network method, the modified method can significantly improve the calculation accuracy.

## 1. Introduction

The artificial neural network was used to solve ordinary differential equations and partial differential equations in the 1990s (Lagaris et al., 1998; Psichogios and Ungar, 1992), but due to the technical development, this research did not attract enough attention. With the explosive growth of data and computing resources, machine learning represented by deep learning has made revolutionary achievements in many fields in recent years, including image recognition (Hafiz et al., 2020), natural language processing (China Bhanja et al., 2019; Pandey et al., 2021), face recognition (Boussaad and Boucetta, 2020), etc. The success of such technology is inseparable from rich training data. Recently, Raissi et al. (2019) promoted this research, extended a set of deep learning methods based on the original one, named it "physics-informed neural network (PINN) ," and used it to solve the forward (Raissi et al., 2017a) and inverse problems (Raissi et al., 2017b) of nonlinear physical models (NPMs). They hope to better apply the PINN method to the modeling and calculation in the field of mathematical physics and engineering, which has triggered a lot of follow-up work, making PINN gradually becomes a research hotspot in this field (Raissi et al., 2020; Wang and Yan, 2021). From a mathematical point of view, a neural network can be regarded as a general nonlinear function approximator, the modeling process of a partial differential equation is also to find the solution satisfying the constraints. Therefore, we can integrate physical laws described by the physical model into the loss function of neural network, to obtain the neural network method with physical law constraints.

PINN has explored many traditional NPM tasks. For example, Fang et al. (2021) studied different soliton solutions of higher-order nonlinear Schrödinger equation (NLSE). Chen et al.

---

[*] Corresponding author email：wangyy424@163.com (Y.Y. Wang); dcq424@126.com (C.Q. Dai)

applied the PINN method to study the propagation of solitons in water based on the KdV equation (Li and Chen, 2020), and the propagation of solitons based on the standard NLSE (Pu et al., 2021). Yan et al. solved the forward and inverse problems of NLSE with the PT-symmetric harmonic potential (Zhou and Yan, 2021) and also discussed the data-driven rogue wave solutions of defocusing NLSE (Wang and Yan, 2021). Since the PINN method was first proposed, researchers have made many improvements to this method in recent years. Meng et al. proposed a parallel PINN method (Meng et al., 2020), which decomposed the long-time problem into multiple short-time problems through a parallel network for parallel calculation and solution, thus solving the problem of excessive training data and greatly accelerating the training speed of the PINN method. Jagtap et al. proposed the conservative PINN method PINN, which decomposes the original solution region into multiple sub-regions, and adds the constraint condition of interface flux conservation in adjacent sub-regions in the loss function (Jagtap et al., 2020). Compared with the original PINN method, it can more accurately simulate the situation with poor smoothness. Lin et al. devised a two-stage PINN method which is tailored to the nature of equations by introducing features of physical systems into neural networks. In stage two, they additionally introduce the measurement of conserved quantities into mean squared error loss to train neural networks (Lin and Chen, 2021). In addition, many other improved PINNs have emerged, such as variational PINN (Kharazmi et al., 2021) and fractional PINN (Mehta et al., 2019; Pang et al., 2019) methods. Meanwhile, many researchers have applied PINN to fluid mechanics (Jin et al., 2021), material mechanics (Chen et al., 2020; Niaki et al., 2021), and other fields.

Among the above improved methods for the PINN method, only the expansion and optimization of the method itself were focused on, and yet the advantages brought by the integrability of NPMs to the expansion and optimization of the PINN method are not paid attention. We know that an essential property of the nonlinear integrable physical model is conservation law and corresponding conserved quantity (Chai et al., 2015). The NLSE is the main physical model describing nonlinear optics. It describes an infinite-dimensional integrable Hamiltonian system, the corresponding has an infinite number of conserved quantities (the concept of conserved quantity here is that a physical quantity of the light field does not change with the change of transmission distance), among which the lowest order conserved quantities are power, energy and momentum, they correspond to different conservation laws (Akhmediev, 1998). These conserved quantities play an important role in studying optical transmissions, such as studying the stability of spatial solitons, momentum exchange in soliton collisions and testing the stability of numerical methods (Zhang et al., 2009). Therefore, it is worthwhile to use these conservation laws and other physical properties of NLSE to further improve the neural network method to study the dynamics behavior of soliton propagation in optical fibers.

The PINN method puts the physical model residual and initial-boundary value residual into the loss function as constraints. Finally, the weight parameters of neural network are obtained by the gradient descent method. We hope to introduce more integrable properties related to physical models, such as conservation laws, into neural networks to characterize the physical properties of physical models and constrain the physical models from a global perspective (Liu et al., 2015). The conservation law is one of the most important physical properties of an integrable model. We consider adding conservation law from the perspective of loss function, in theory, which can bring strong binding force for a neural network to solve the physical model. Therefore, a modified PINN method based on the conservation law constraint is proposed in this paper. In other words, the conservation law of NLSE is added to the loss function to obtain the conservation law residual as the constraint. In this paper, we integrate the conservation of momentum and energy into the design of loss function of neural network, to obtain the neural network with physical property constraints. The trained network can not only better approximate the observed data, but also satisfy the conservation property followed by the NPMs.

The modified PINN method based on the conservation law constraint has the following advantages. (i) Strong restraint. After adding the physical law followed by the equations, the constraint effect on the training results of neural network is better; (ii) wide range of application. It has good effects for different optical solitons; and (iii) the small training error. Compared with the classical PINN method, the calculation accuracy of the improved method is significantly improved.

## 2. The modified PINN method based on the conservation law constraints

In this paper, the conservation law is used to constrain the training process of neural networks to reconstruct the dynamic characteristics and parameters of NLSE. The general form of NLSE is considered

$$Q_z + N(\lambda, Q, Q_t, Q_{ttt}, \cdots,) = 0, \quad z \in (z_1, z_2), \; t \in (t_1, t_2), \tag{1}$$

where $N$ is a linear and nonlinear differential operator, $Q = r + im$. PINN considers establishing a neural network to approximate the function $Q$, in fact, we need to approximate the real part $r$ and imaginary part $m$ of the function, respectively.

$$\begin{aligned} f_r &:= r_z + N_r(\lambda, r, r_t, r_{ttt}, \cdots,) \\ f_m &:= m_z + N_m(\lambda, m, m_t, m_{ttt}, \cdots,). \end{aligned} \tag{2}$$

Next, we give the loss function of different conservation law constraints

$$L_1 = MSE_e + MSE_{bc} + MSE_{ic}, \tag{3}$$

$$L_2 = MSE_e + MSE_{en} + MSE_{bc} + MSE_{ic}, \tag{4}$$

$$L_3 = MSE_m + MSE_{en} + MSE_{bc} + MSE_{ic}, \tag{5}$$

$$L_4 = MSE_m + MSE_{bc} + MSE_{ic}, \tag{6}$$

where $L_1$ is the loss function of classical PINN method, $L_2, L_3$ and $L_4$ are the loss functions of different conservation law combinations. In the following article, we will discuss the advantages of PINN with conservation constraints. $MSE$ represents various mean square errors

$$\begin{aligned} MSE_{ic} &= \frac{1}{N_0} \sum_{i=1}^{N_0} (|r(z^i, t^i) - r^i|^2 + |m(z^i, t^i) - m^i|^2), \\ MSE_{bc} &= \frac{1}{N_b} \sum_{q=1}^{N_b} (|r_1(z^q, t^q) - r_1^q|^2 + |m_1(z^q, t^q) - m_1^q|^2), \\ MSE_e &= \frac{1}{N_f} \sum_{j=1}^{N_f} (|f_{er}(z^j, t^j)|^2 + |f_{em}(z^j, t^j)|^2), \\ MSE_m &= \frac{1}{N_f} \sum_{j=1}^{N_f} (|f_{mr}(z^j, t^j)|^2 + |f_{mm}(z^j, t^j)|^2), \\ MSE_{en} &= \frac{1}{N_f} \sum_{j=1}^{N_f} (|f_{enr}(z^j, t^j)|^2 + |f_{enm}(z^j, t^j)|^2), \end{aligned} \tag{7}$$

where $MSE_{ic}$ and $MSE_{bc}$ represent the training data obtained from the initial state and boundary state, $MSE_e$ represents the residual obtained by the equation model, $MSE_m$ is the residual obtained from momentum conservation, and $MSE_{en}$ is the residual obtained from energy conservation. The neural network learns the parameters such as weights and biases by minimizing the mean square error of the loss function. In this paper, $N_0 = 50, N_b = 50, N_f = 10000$. The neural network has 6 layers, with 50 neurons in each layer.

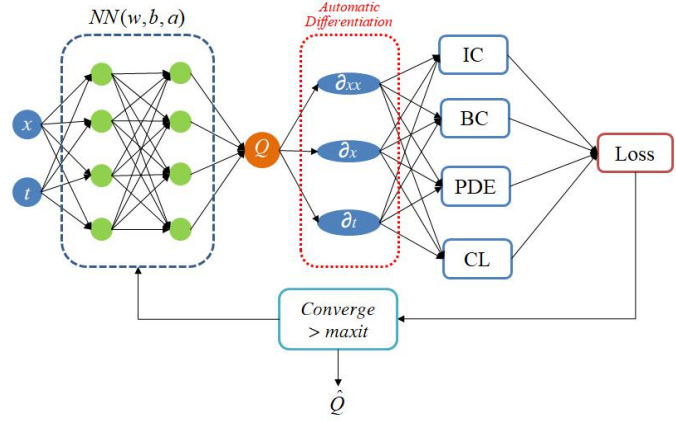

**Fig. 1.** Schematic diagram of the modified PINN method based on the conservation law constraints for NLSE model. In addition to initial-boundary conditions and partial differential equations, the method adds conservation law conditions. Here IC and BC respectively mean initial and boundary conditions, PED denotes partial differential equation, and CL and $\hat{Q}$ respectively represent conservation law and prediction results.

### 3. Data-driven optical solitons

In optics, the propagation of nonlinear waves in optical fiber can be described by standard NLSE (Agrawal, 2000)

$$iQ_z + Q_{tt} + 2|Q|^2 Q = 0, \tag{8}$$

where $Q$ is the pulse slowly varying amplitude envelope, $z,t$ represents the normalized distance and time coordinate of optical solitons propagating along with the fiber in a single-mode fiber. Eq. (1) describes the evolution of nonlinear waves in single-mode fiber, and the nonlinear term is caused by self-phase modulation in the fiber, the combined effect of the nonlinear term and group velocity dispersion produces optical solitons. We also give the conservation laws of momentum and energy for Eq. (1) (Zhang et al., 2009)

$$\text{Conservation of energy:} \quad (|Q|^2)_z - i(Q_t Q^* - Q Q_t^*)_t = 0, \tag{9}$$

$$\text{Conservation of momentum:} \quad (Q Q_t^*)_z - i(Q_t Q_t^* - Q Q_{tt}^* - |Q|^4)_t = 0, \tag{10}$$

where $Q^*$ is the conjugate of the amplitude envelope $Q$.

Next, we discuss four different solitons in single-mode fiber transmission, including one-soliton, soliton molecules, interaction of two-soliton, and rogue wave by using the modified PINN method based on the conservation law constraints.

### 3.1. One-soliton

The exact one-soliton solution reads (Pu et al., 2021)

$$Q=0.6\,\text{sech}(0.6t)e^{0.36iz+i}, \quad t \in [-15,15], z \in [0,5]. \tag{11}$$

From the known range of the space-time region, the initial and boundary conditions of the data can be derived, and the data sets can also be obtained by the pseudo-spectral method, data points are obtained by discrete the exact one-soliton. In this paper, the size of the data set obtained by the numerical method is $[256 \times 201]$.

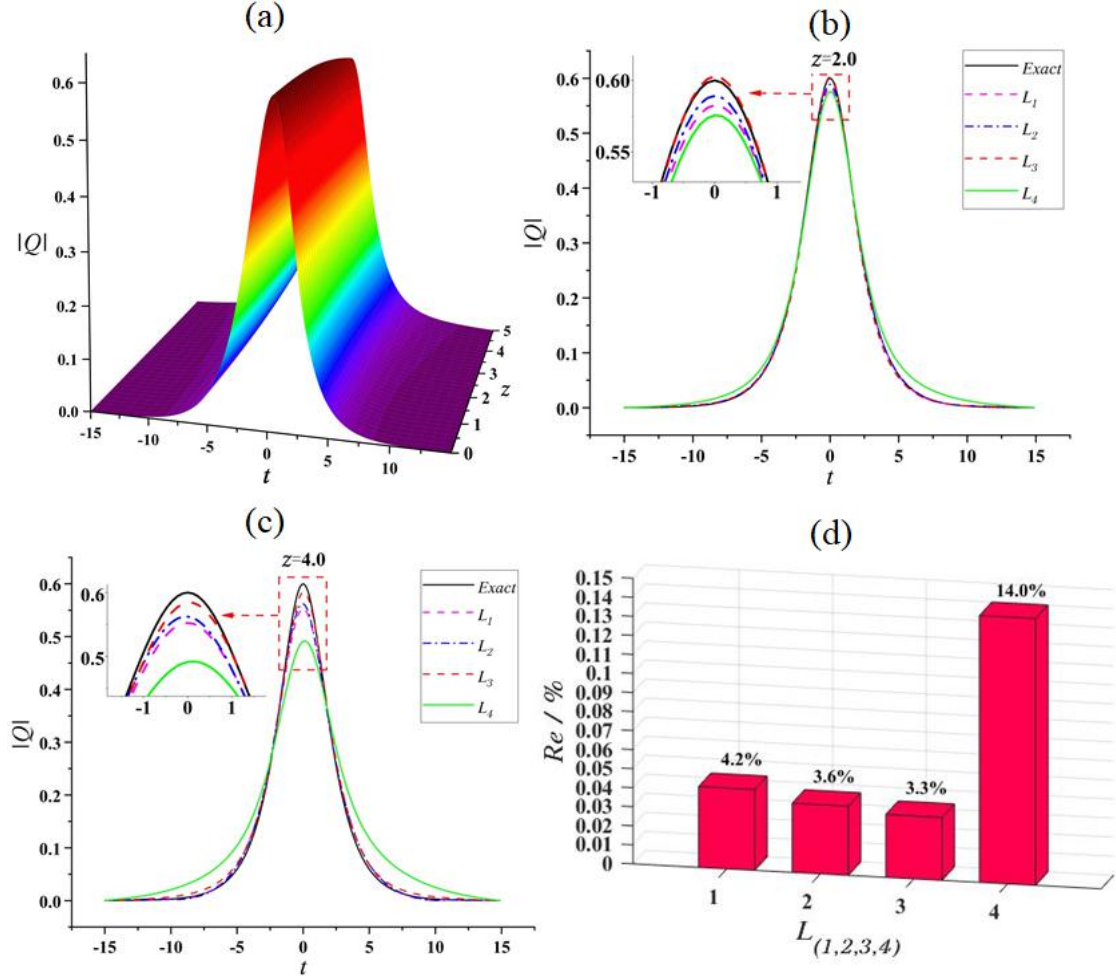

**Fig. 2.** Prediction results of one-soliton: (a) 3D diagram of one-soliton prediction solution under $L_3$ condition; (b) Comparison diagram of prediction solution and exact solution at $z = 2.0$; (c) Comparison diagram of prediction solution and exact solution at $z = 4.0$; (d) The relative percentages errors of the prediction solution under the condition of $L_1, L_2, L_3, L_4$.

Fig. 2(a) shows the 3D diagram of the prediction results of the propagation process of one-soliton under the $L_3$ constraint. Pu et al. used the classical PINN method to solve one-soliton (Pu et al., 2021). Compared with it, our prediction distance is greatly improved. One of the properties of solitons is that they can transmit stably along the current direction and keep the amplitude unchanged. Figs. 2(b)-(c) show the comparison between the predicted value and the exact value at $z = 2.0$ and $z = 4.0$. From the comparison results, it can be seen that compared with the classical PINN method $L_1$, the predicted value of $L_2, L_3$ are closer to the exact value, and the stable transmission is maintained for a longer distance. Meanwhile, after a long-distance transmission, there has been a large error in the amplitude of $L_4$, which violates the transmission properties of solitons. Fig. 2 (d) shows the relative errors $Re(Re = (|\hat{Q} - Q|/Q) \times 100\%)$ in four cases. It can be seen from the results that the relative errors of $L_2$ and $L_3$ are smaller, which shows that the constraint effect of both is better and the error of prediction results is smaller. In summary, from multiple perspectives, in the prediction of one-soliton, $L_2$ and $L_3$ (combined constraints of equation and energy conservation and combined constraints of momentum and energy conservation) have obvious advantages compared with the classical PINN method. In contrast, the single momentum conservation constraint in $L_4$ cannot provide good constraint effect.

### 3.2. Soliton molecules

The exact soliton molecules solution reads (Wang, B. et al., 2020)

$$Q = \frac{-2i(-0.05ie^{0.7744iz-0.8t} - 0.06ie^{0.7744iz+0.8t} - 0.05ie^{0.64iz+0.88t} - 0.05ie^{0.64iz-0.88t})}{-1.32e^{-0.1344iz} - 1.32e^{0.1344iz} - 1.41e^{-0.08t} - 1.23e^{0.08t} + 0.0004e^{-1.68t} + 0.0037e^{1.68t}}, \quad (12)$$

with the space-time region $t \in [-15, 15], z \in [0, 3]$.

Fig. 3 (a) shows the propagation process of the predicted soliton molecules, which maintains stable propagation at a certain distance. The two-solitons achieve the velocity resonance to form a bound state and keep an equal distance parallelly transmit without any interaction, and their amplitudes remain unchanged. In Fig. 3 (b), the relative errors between the prediction results and the exact solution under four conditions are given. The results show that compared with the classical PINN method, $L_2$ and $L_3$ have obvious advantages, and the value of relative error is significantly reduced, while $L_4$ method also has no advantages in the prediction of soliton molecules. Fig. 3 (c) provides the convergence curves of the loss function with the number of iterations under four conditions. It can be seen that the final convergence errors of $L_2$ and $L_3$ are smaller than $L_1$, which proves the excellent performance of $L_2$ and $L_3$ again. Fig. 3 (d)-(f) respectively show the density plots of the absolute errors $Er(Er = \hat{Q} - Q)$ of $L_1, L_2$ and $L_3$. From the error values and chromaticity of the density plots, it can be seen that with the increasing of the prediction distance, the error of the classical PINN method increases faster and the absolute error is larger. From the above analysis, it can be seen that $L_2$ and $L_3$ have more advantages than $L_1$. In other words, the PINN method with the combined constraints of equation and energy conservation and the combined constraints of momentum and energy conservation has more obvious prediction advantages than the traditional PINN method. Similarly to the prediction of one-soliton, the prediction result of soliton molecules is also excellent and achieves the desired effects.

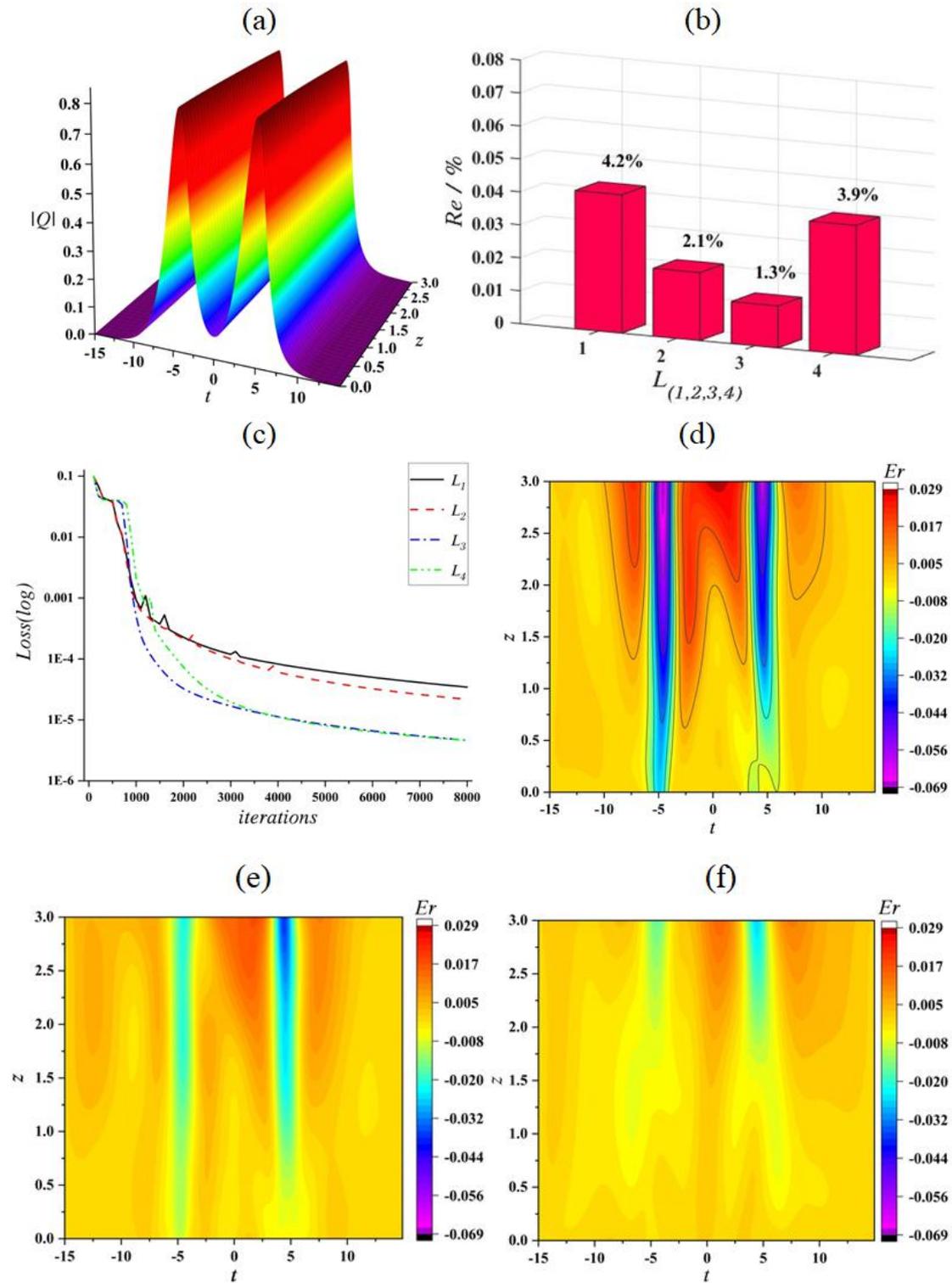

**Fig. 3.** Prediction results of soliton molecules: (a) 3D diagram of soliton molecules prediction solution under $L_3$ condition; (b) The relative percentages errors of the prediction solution under the condition of $L_1, L_2, L_3, L_4$ ; (c) Convergence curve of loss function of $L_1, L_2, L_3$ and $L_4$ ; The absolute error density diagram of the prediction result of (d) $L_1$, (e) $L_2$, (f) $L_3$ and exact solutions.

### 3.3. Interaction of two-soliton

The exact solution of the two-soliton interaction is as follows (Pu et al., 2021)

$$Q = \frac{-2i(-0.246ie^{0.64iz+1.4t} + 0.462ie^{1.96iz-0.8t} - 0.264ie^{0.64iz-1.4t} + 0.462ie^{1.96iz+0.8t})}{-1.12e^{-1.32iz} - 1.12e^{1.32iz} + 1.21e^{-0.6t} + 1.21e^{0.6t} + 0.09e^{-2.2t} + 0.09e^{2.2t}}, \quad (13)$$

with the space-time region $t \in [-6,6], z \in [-2,2]$.

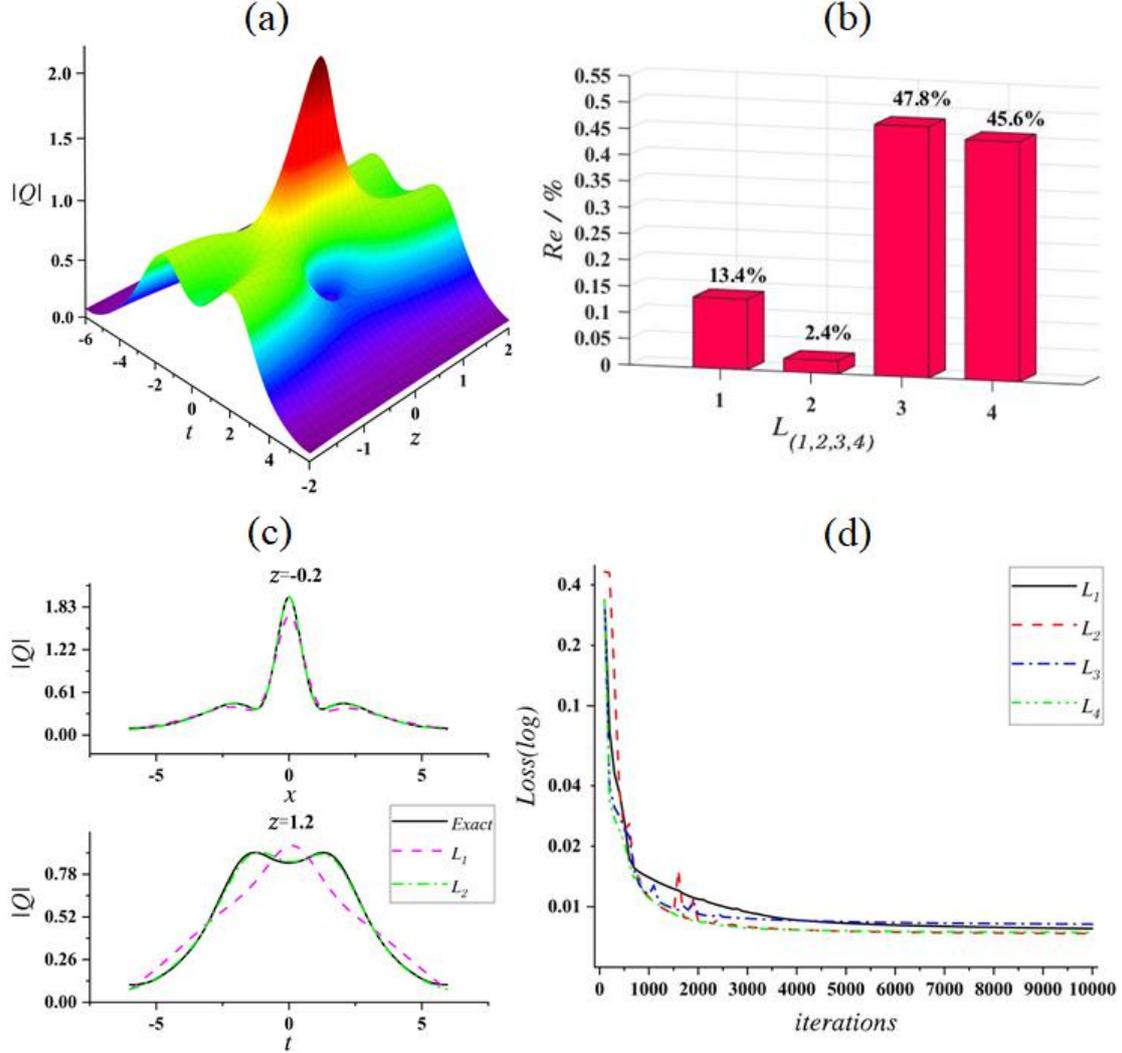

**Fig. 4**. Prediction results of two-soliton interaction: (a) 3D diagram of predicted results of two-soliton interaction under $L_2$ condition; (b) The relative percentages errors of the prediction solution under the condition of $L_1, L_2, L_3, L_4$; (c) Comparison of exact solutions and predicted results at different propagation distances; (d) Convergence curve of loss function of $L_1, L_2, L_3$ and $L_4$.

In Fig. 4(a), the dynamic characteristic diagram of the predicted two-soliton interaction under the condition of $L_2$ is shown, the predicted results conform to the properties of solitons. Elastic scattering occurs when the two solitons interact with each other. That is, the original direction and amplitude are still maintained after the interaction, the waveform and wave velocity can return to the original state, and there is no energy transfer between them. Meanwhile, the study of using classical PINN to predict the interaction between solitons can be obtained in (Pu et al., 2021). The relative errors in four different cases are given in Fig. 4 (b). It can be seen that the relative error of the prediction result under $L_2$ condition has reached an ideal result. In contrast, $L_3$, which has great advantages in both one-soliton and soliton molecules, has a great error. The results show that neither $L_3$ nor $L_4$ has an advantage in the prediction of soliton interaction. In Fig. 4 (c), we compare the exact solution and predicted solution of two-soliton interaction at $z = -0.2$ and $z = 1.2$, the black solid line is the exact solution, and the red and green dotted lines are the predicted results under $L_1$ and $L_2$ conditions, respectively. It can be seen that the coincidence degree of the green dotted line and the exact solution is higher and the error is smaller, which is consistent with the results in Fig. 4 (b). Fig. 4 (d) provides the convergence curves of the loss function with the number of iterations under four conditions. It can be seen that under the same number of iterations, the loss function of $L_2$ is the smallest and converges to $7.347 \times 10^{-3}$ after 10000 iterations. From the comparison of several aspects in Fig. 4, it can be concluded that the prediction result under the condition of $L_2$ is the best, that is, the PINN method combined with equation and energy conservation constraints has more obvious prediction advantages than the traditional PINN method.

### 3.4. Rogue wave

The exact solution of the rogue wave is as follows (Pu et al., 2021)

$$Q = 0.6e^{0.72iz}(1 - \frac{4+5.76iz}{1+1.44t^2+2.0736z^2}), \quad t \in [-3,3], z \in [-1.5,1.5]. \tag{14}$$

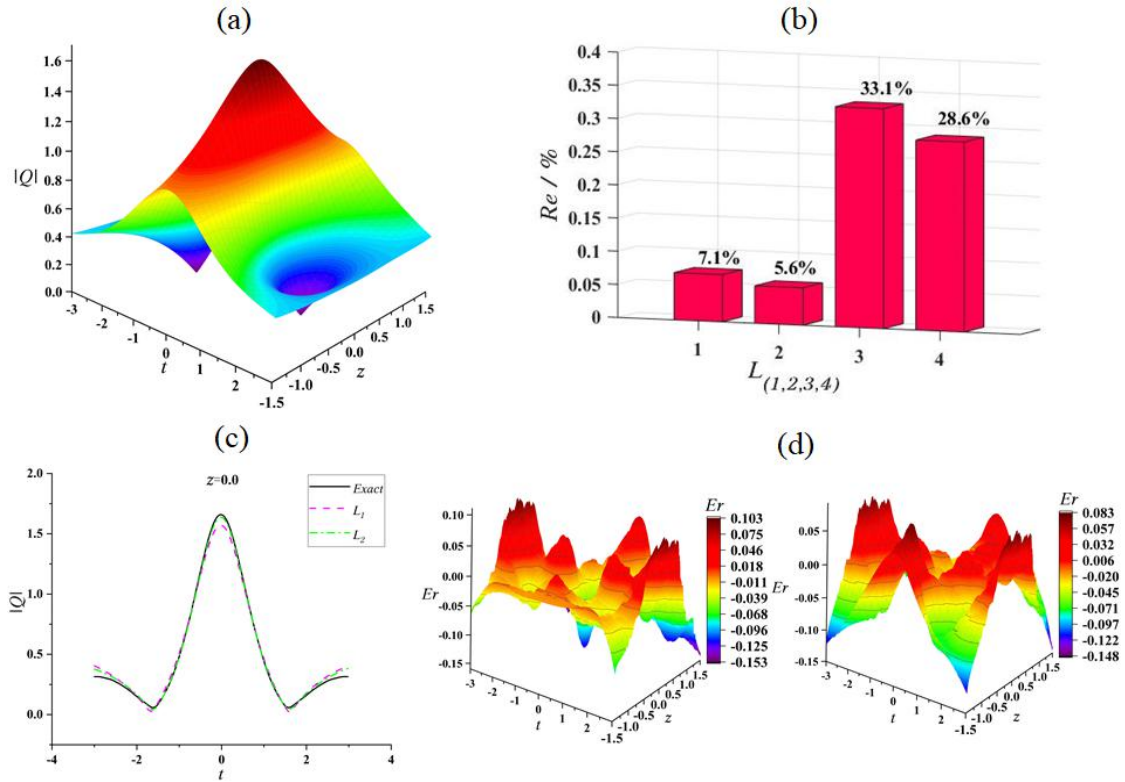

**Fig. 5.** Prediction results of rogue wave: (a) 3D diagram of rogue wave prediction solution under $L_2$ condition; (b) The relative percentages errors of the prediction solution under the condition of $L_1, L_2, L_3, L_4$; (c) Comparison diagram of prediction solution and exact solution at $z = 0.0$; (d) The absolute error 3D diagram of the prediction results and exact solutions, with the result of $L_1$ on the left and the result of $L_2$ on the right.

In Fig. 5(a), the predicted rogue wave exhibits high amplitude in the optical fiber. Due to the modulation instability in the optical fiber, the weak modulation on the plane wave can produce exponential growth along the transmission distance, and result in rogue waves. In Fig. 5(b), the relative errors under the four conditions are given, which are similar to the results of the interaction of two-soliton, and only $L_2$ has an advantage over the classical PINN method. In Fig. 5 (c), we compare the exact solution of the strange wave at $z = 0.0$ with the prediction results of $L_1$ and $L_2$, and it can be seen from the figure that the coincidence effect of $L_2$ is better. Fig. 5 (c) shows the absolute error 3D diagram of the exact solution and the predicted solution, in which the left is the error diagram of $L_1$ and the right is the error diagram of $L_2$. From the error scale value, it can be concluded that the error of $L_2$ is smaller. In conclusion, the prediction effect of rogue wave and two-soliton interaction is similar, and the prediction result of $L_2$ is the best. That is, the PINN method combined with equation and energy conservation constraints has more obvious prediction advantages than the traditional PINN method.

## 4. Parameter prediction of physical model

In this section, we will consider the parameter discovery problem of a data-driven NLSE model. The equation is as follows

$$iQ_z + \lambda_1 Q_{tt} + \lambda_2 |Q|^2 Q = 0, \qquad (15)$$

where slowly varying envelope $Q$ contains real part $r$ and imaginary part $m$, and variables $\lambda_1, \lambda_2$ are the unknown dispersion and nonlinear coefficient to be trained.

The physical model is transformed into

$$\begin{aligned} f_r &:= r_z + \lambda_1 m_{tt} + \lambda_2 (r^2 + m^2)m, \\ f_m &:= m_z - \lambda_1 r_{tt} - \lambda_2 (r^2 + m^2)r. \end{aligned} \qquad (16)$$

In the inverse problem, we obtain the approximate value of these unknown coefficients by minimizing the loss functions $L_1$ and $L_2$. Meanwhile, the rogue wave solution is taken as the data set, and the size of the data set and the structure of the neural network have been given above.

Table.1. Comparison of Correct equations and identified equations obtained by PINN

| Item NLSE | Category of data sets | Loss | Equation | Relative error $Re$ | |
|---|---|---|---|---|---|
| | | | | $\lambda_1$ | $\lambda_2$ |
| Correct equations | | | $iQ_z + Q_{tt} + 2|Q|^2 Q = 0$ | | |
| Identified equations | one soliton | $L_1$ | $iQ_z + 0.997709 Q_{tt} + 1.994775 |Q|^2 Q = 0$ | 0.17901% | 0.22398% |
| | | $L_2$ | $iQ_z + 0.997771 Q_{tt} + 1.994963 |Q|^2 Q = 0$ | 0.17465% | 0.21675% |
| | two soliton molecule | $L_1$ | $iQ_z + 0.997914 Q_{tt} + 1.995404 |Q|^2 Q = 0$ | 0.23837% | 0.24446% |
| | | $L_2$ | $iQ_z + 0.997964 Q_{tt} + 1.995671 |Q|^2 Q = 0$ | 0.19672% | 0.21740% |
| | two soliton interaction | $L_1$ | $iQ_z + 1.001130 Q_{tt} + 2.001166 |Q|^2 Q = 0$ | 0.09093% | 0.07092% |

| | | $L_2$ | $iQ_z + 0.999588Q_{tt} + 2.000491|Q|^2 Q = 0$ | 0.06780% | 0.02700% |
| | rogue wave | $L_1$ | $iQ_z + 1.024685Q_{tt} + 2.015894|Q|^2 Q = 0$ | 2.53365% | 0.78839% |
| | | $L_2$ | $iQ_z + 1.013357Q_{tt} + 2.008907|Q|^2 Q = 0$ | 1.38358% | 0.42123% |

In Table. 1, we take four different types of solitons as data sets to observe the training results and corresponding errors of unknown coefficients in $L_1$ and $L_2$ cases. It can be seen from the value of relative error that $L_2$ has obvious advantages over $L_1$ in the four data sets, that is, the PINN method combined with equation and energy conservation constraints has more obvious prediction advantages than the traditional PINN method. In addition, it can be concluded through comparison that the prediction accuracy of $L_2$ is improved most obviously when the rogue wave solution is used as the data set. The results show that the conservation law constraint is still effective in the inverse problem.

## 5. Conclusion

In this paper, we introduce the modified PINN method based on the conservation law constraint. We hope to introduce the conservation laws into the neural network and design a more targeted PINN method, which requires mining the underlying information of the given equation to improve accuracy and reliability. Therefore, we propose to add conservation law constraints to the loss function, and apply the modified PINN method to the NLSE to verify the feasibility.

Modifying the classical PINN method, we propose three forms of loss function with conservation law constraint, and discuss the advantages and disadvantages of solving different soliton solutions. Compared with the classical PINN method, the results of one-soliton and soliton molecules are similar, $L_2$ and $L_3$ can obviously improve the prediction accuracy, while the constraint effect of $L_4$ is not ideal. In other words, the PINN method with the combined constraints of equation and energy conservation and the combined constraints of momentum and energy conservation has more obvious prediction advantages than the traditional PINN method. For two-soliton interaction and rogue wave, adding equation and energy conservation constraint have obvious advantages. In the parameter prediction of the physical model, $L_2$ still has an obvious advantage through the comparison of relative errors. Meanwhile, it can maintain the advantage for different solitons as data sets, the PINN method combined with equation and energy conservation constraints has more obvious prediction advantages than the traditional PINN method. In addition, by analyzing the results of four data sets, it can be concluded that the prediction accuracy of $L_2$ is improved most obviously when the rogue wave solution is used as the data set. By verifying the forward and inverse problems of the physical model, the results show that our improvement is of great significance for predicting NPMs.

Our research can be more targeted to solve NPMs and promote the development of this field. However, our neural network method increases learning costs and training time. Next, we will continue to improve the method and improve the accuracy without reducing the learning efficiency. Meanwhile, we will try to extend this method to other models to improve the adaptability and generalization ability.

**Acknowledgements**

This work is supported by the Zhejiang Provincial Natural Science Foundation of China (Grant No. LR20A050001), the National Natural Science Foundation of China (Grant Nos. 11874324, 12075210) and Scientific Research and Developed Fund of Zhejiang A&F University (Grant No. 2021FR0009)

**Conflict of interest**

The authors have declared that no conflict of interest exists.

**Ethical Standards**

This Research does not involve Human Participants and/or Animals.

**References**

Agrawal, G. P. (2000). Nonlinear fiber optics. In Nonlinear Science at the Dawn of the 21st Century, Springer.

Akhmediev, N. N., 1998. Spatial solitons in Kerr and Kerr-like media. Optical and Quantum Electronics, 30, 535-569. https://doi:10.1023/A:1006902715737

Boussaad, L., Boucetta, A., 2020. Deep-learning based descriptors in application to aging problem in face recognition. Journal of King Saud University - Computer and Information Sciences. https://doi:10.1016/j.jksuci.2020.10.002

Chai, J., Tian, B., Zhen, H. L., Sun, W. R., 2015. Conservation laws, bilinear forms and solitons for a fifth-order nonlinear Schrödinger equation for the attosecond pulses in an optical fiber. Annals of Physics, 359, 371-384. https://doi:10.1016/j.aop.2015.04.010

Chen, Y., Lu, L., Karniadakis, G. E., Dal Negro, L., 2020. Physics-informed neural networks for inverse problems in nano-optics and metamaterials. Opt Express, 28, 11618-11633. https://doi:10.1364/OE.384875

China Bhanja, C., Laskar, M. A., Laskar, R. H., Bandyopadhyay, S., 2019. Deep neural network based two-stage Indian language identification system using glottal closure instants as anchor points. *Journal of King Saud University - Computer and Information Sciences*. https://doi:10.1016/j.jksuci.2019.07.001

Fang, Y., Wu, G. Z., Wang, Y. Y., Dai, C. Q., 2021. Data-driven femtosecond optical soliton excitations and parameters discovery of the high-order NLSE using the PINN. Nonlinear Dynamics, 105, 603-616. https://doi:10.1007/s11071-021-06550-9

Hafiz, R., Haque, M. R., Rakshit, A., Uddin, M. S., 2020. Image-based soft drink type classification and dietary assessment system using deep convolutional neural network with transfer learning. Journal of King Saud University - Computer and Information Sciences. https://doi:10.1016/j.jksuci.2020.08.015

Jagtap, A. D., Kharazmi, E., Karniadakis, G. E., 2020. Conservative physics-informed neural networks on discrete domains for conservation laws: Applications to forward and inverse problems. Computer Methods in Applied Mechanics and Engineering, 365, 113028. https://doi:10.1016/j.cma.2020.113028

Jin, X. W., Cai, S. Z., Li, H., Karniadakis, G. E., 2021. NSFnets (Navier-Stokes flow nets): Physics-informed neural networks for the incompressible Navier-Stokes equations. Journal of

Computational Physics, 426, 109951. https://doi:10.1016/j.jcp.2020.109951

Kharazmi, E., Zhang, Z., Karniadakis, G. E. M., 2021. hp-VPINNs: Variational physics-informed neural networks with domain decomposition. Computer Methods in Applied Mechanics and Engineering, 374, 113547. https://doi:10.1016/j.cma.2020.113547

Lagaris, I. E., Likas, A., Fotiadis, D. I., 1998. Artificial Neural Networks for Solving Ordinary and Partial Differential Equations. IEEE Transactions on Neural Networks, 9, 987-1000. https://doi:10.1109/72.712178

Li, J., Chen, Y., 2020. A deep learning method for solving third-order nonlinear evolution equations. Communications in Theoretical Physics, 72, 115003. https://doi:10.1088/1572-9494/abb7c8

Liu, D. Y., Tian, B., Sun, W. R., Wang, Y. P., 2015. Conservation laws and dark-soliton solutions of an integrable higher-order nonlinear Schrödinger equation for a density-modulated quantum condensate. Physica Scripta, 90, 045205. https://doi:10.1088/0031-8949/90/4/045205

Li, S. N., Chen, Y., 2021. A two-stage physics-informed neural network method based on conserved quantities and applications in localized wave solutions. arXiv:2107.01009v1 [nlin.SI] 2 Jul 2021

Mehta, P. P., Pang, G., Song, F., Karniadakis, G. E., 2019. Discovering a universal variable-order fractional model for turbulent Couette flow using a physics-informed neural network. Fractional Calculus and Applied Analysis, 22, 1675-1688. https://doi:10.1515/fca-2019-0086

Meng, X. H., Li, Z., Zhang, D. K., Karniadakis, G. E., 2020. PPINN: Parareal physics-informed neural network for time-dependent PDEs. Computer Methods in Applied Mechanics and Engineering, 370, 113250. https://doi:10.1016/j.cma.2020.113250

Niaki, S. A., Haghighat, E., Campbell, T., Poursartip, A., Vaziri, R., 2021. Physics-informed neural network for modelling the thermochemical curing process of composite-tool systems during manufacture. Computer Methods in Applied Mechanics and Engineering, 384, 113959. https://doi:10.1016/j.cma.2021.113959

Pandey, B., Kumar Pandey, D., Pratap Mishra, B., Rhmann, W., 2021. A comprehensive survey of deep learning in the field of medical imaging and medical natural language processing: Challenges and research directions. *Journal of King Saud University - Computer and Information Sciences*. https://doi:10.1016/j.jksuci.2021.01.007

Pang, G. F., Lu, L., Karniadakis, G. E. M., 2019. fPINNs: Fractional Physics-Informed Neural Networks. SIAM Journal on Scientific Computing, 41, A2603-A2626. https://doi:10.1137/18m1229845

Psichogios, D. C., Ungar, L. H., 1992. A Hybrid Neural Network-First Principles Approach to Process Modeling. AIChE Journal, 38, 1499-1511. https://doi:10.1002/aic.690381003

Pu, J. C., Li, J., Chen, Y., 2021. Soliton, breather, and rogue wave solutions for solving the nonlinear Schrödinger equation using a deep learning method with physical constraints. Chinese Physics B, 30, 060202. https://doi:10.1088/1674-1056/abd7e3

Raissi, M., Perdikaris, P., Karniadakis, G. E., 2017a. Physics Informed Deep Learning (Part I): Data-driven Solutions of Nonlinear Partial Differential Equations. arXiv:1711.10561v1 [cs.AI] 28 Nov 2017

Raissi, M., Perdikaris, P., Karniadakis, G. E., 2017b. Physics Informed Deep Learning (Part II): Data-driven Discovery of Nonlinear Partial Differential Equations. arXiv:1711.10566v1 [cs.AI] 28 Nov 2017

Raissi, M., Perdikaris, P., Karniadakis, G. E., 2019. Physics-informed neural networks: A deep learning framework for solving forward and inverse problems involving nonlinear partial differential equations. Journal of Computational Physics, 378, 686-707. https://doi:10.1016/j.jcp.2018.10.045

Raissi, M., Yazdani, A., Karniadakis, G. E., 2020. Hidden fluid mechanics: Learning velocity and pressure fields from flow visualizations. Science, 367, 1026-1030. https://doi:10.1126/science.aaw4741

Wang, B., Zhang, Z., Li, B., 2020. Soliton Molecules and Some Hybrid Solutions for the Nonlinear Schrödinger Equation. Chinese Physics Letters, 37, 030501. https://doi:10.1088/0256-307x/37/3/030501

Wang, L., Yan, Z. Y., 2021. Data-driven rogue waves and parameter discovery in the defocusing nonlinear Schrodinger equation with a potential using the PINN deep learning. Physics Letters A, 404, 127408. https://doi:10.1016/j.physleta.2021.127408

Wang, L., Yan, Z., 2021. Data-driven rogue waves and parameter discovery in the defocusing nonlinear Schrödinger equation with a potential using the PINN deep learning. Physics Letters A, 404, 127408. https://doi:10.1016/j.physleta.2021.127408

Zhang, H. Q., Tian, B., Meng, X. H., Lü, X., Liu, W. J., 2009. Conservation laws, soliton solutions and modulational instability for the higher-order dispersive nonlinear Schrödinger equation. The European Physical Journal B, 72, 233-239. https://doi:10.1140/epjb/e2009-00356-3

Zhou, Z., Yan, Z., 2021. Solving forward and inverse problems of the logarithmic nonlinear Schrödinger equation with PT-symmetric harmonic potential via deep learning. Physics Letters A, 387, 127010. https://doi:10.1016/j.physleta.2020.127010